\documentclass{aa}
\usepackage[utf8]{inputenc}
\usepackage{comment}
\usepackage{amsmath}
\usepackage{graphicx}
\usepackage{gensymb}
\usepackage[colorlinks=true,linkcolor=blue,citecolor=blue]{hyperref}
\usepackage{xspace}

\usepackage{arydshln} %to make dashed lines in Table 2

\newcommand\titlelowercase[1]{\texorpdfstring{\lowercase{#1}}{#1}}
\newcommand{\rproc}{\textit{r}-process\xspace}
\usepackage{natbib}
\bibpunct{(}{)}{;}{a}{}{,} % to follow the A&A style

\begin{document}
\title{ Discovery of a 760\,nm P~Cygni line in AT2017gfo: Identification of yttrium in the kilonova photosphere} 
\titlerunning{An yttrium 760\,nm P~Cygni line in the kilonova AT2017gfo}

\author{Albert Sneppen\inst{\ref{addr:DAWN},\ref{addr:jagtvej}}\thanks{Corresponding author email: albert.sneppen@nbi.ku.dk } \and
Darach Watson\inst{\ref{addr:DAWN},\ref{addr:jagtvej}}
}

\institute{Cosmic Dawn Center (DAWN)\label{addr:DAWN}
\and
Niels Bohr Institute, University of Copenhagen, Blegdamsvej 17, K{\o}benhavn 2100, Denmark\label{addr:jagtvej} 
}

% The list of authors, and the short list which is used in the headers.
% If you need two or more lines of authors, add an extra line using \newauthor
%\author[0000-0002-5460-6126]{Albert Sneppen}
%\affiliation{Cosmic Dawn Center (DAWN)}
%\affiliation{Niels Bohr Institute, University of Copenhagen, Lyngbyvej 2, K\o benhavn \O~2100, Denmark}

%\author{Darach Watson}
%\affiliation{Cosmic Dawn Center (DAWN)}
%\affiliation{Niels Bohr Institute, University of Copenhagen, Lyngbyvej 2, K\o benhavn \O~2100, Denmark}

%\author{Other Co-authors}
\date{Received 15 March 2023 / Accepted 3 June 2023 }

\abstract{
    Neutron star mergers are believed to be a major cosmological source of rapid neutron-capture elements. The kilonovae associated with neutron star mergers have to date yielded only a single well-identified spectral signature: the P~Cygni line of Sr\(^+\) at about \(1\,\mu\)m in the spectra of the optical transient, AT2017gfo. Such P~Cygni lines are important, because they provide significant information not just potentially on the elemental composition of the merger ejecta, but also on the velocity, geometry, and abundance stratification of the explosion. In this paper we show evidence for a previously unrecognised P~Cygni line in the spectra of AT2017gfo that emerges several days after the explosion, located at $\lambda \approx 760\,$nm. We show that the feature is well-reproduced by 4d\(^2\)--4d5p transitions of Y\(^+\), which have a weighted mean wavelength around 760--770\,nm, with the most prominent line at 788.19\,nm. While the observed line is weaker than the Sr\(^+\) feature, the velocity stratification of the new line provides an independent constraint on the expansion rate of the ejecta similar to the constraints from Sr\(^+\). \newline 
}

%Neutron star mergers are believed to be a major cosmological source of rapid neutron-capture elements. The kilonovae associated with neutron star mergers have to date yielded only a single well-identified spectral signature: the P Cygni line of Sr$^+$ at about 1$\mu$m in the spectra of the optical transient, AT2017gfo. Such P Cygni lines are important, because they provide significant information not just potentially on the elemental composition of the merger ejecta, but also on the velocity, geometry, and abundance stratification of the explosion. In this paper we show evidence for a previously unrecognised P Cygni line in the spectra of AT2017gfo that emerges several days after the explosion, located at $\lambda \approx 760\,$nm. We show that the feature is well-reproduced by 4d$^2$-4d5p transitions of Y$^+$, which have a weighted mean wavelength around 760-770 nm, with the most prominent line at 788.19 nm. While the observed line is weaker than the Sr$^+$ feature, the velocity stratification of the new line provides an independent constraint on the expansion rate of the ejecta similar to the constraints from Sr$^+$.

\maketitle

\section{Introduction}
Neutron star mergers are a significant source of \rproc elements, as suggested by \cite{Lattimer1977} and \cite{Symbalisty1982} and indicated by observations of kilonovae \citep[e.g.][]{Tanvir2013,Coulter2017}, and with confirmation provided by spectroscopy \citep{Watson2019}. %Determining the abundance and elementary distribution is essential for understanding spectral synthesis in the Universe and the sites producing \rproc elements.

The most detailed constraints on any kilonova so far are provided by the series of X-shooter spectra for AT2017gfo \citep{Smartt2017,Pian2017}. These daily spectra taken from 1.4 days and onwards after the event provide the temporal evolution of the transient in continuum, emission, and absorption features from the ultraviolet to the near-infrared. It was these data that yielded the first observational identification of a freshly synthesised \rproc element in an astrophysical site \citep{Watson2019}. This identification came in the form of resonance lines of Sr\(^+\) and demonstrated that the early spectra of AT2017gfo can be explained by lighter \rproc material. These findings were reproduced and extended with systematic analyses using radiative transfer codes with improved atomic line data \citep{Domoto2021,Gillanders2022,Vieira2023}. Another proposed interpretation is that this P~Cygni feature originates from the helium 1083.3\,nm line, which, as shown in \citet{Perego2022} and \citet{Tarumi2023}, in non-local-thermodynamic-equilibrium (NLTE) conditions could reproduce the observed feature. In addition to Sr\(^+\), the potential presence of Zr\(^+\) \citep{Gillanders2022}, Y\(^+\), and Zr\(^+\) \citep{Vieira2023} has been inferred at days 1--2 post-merger using radiative transfer modelling, principally based on absorption features at wavelengths $\lambda \leq500$\,nm. The identification of features at $\lambda \approx$ 1200--1400\,nm with absorption due to La\(^{2+}\) and Ce\(^{2+}\) has been proposed \citep{Domoto2022}. On the other hand, \citet{Watson2019} noted several near-infrared emission features that emerge most strongly several days post-merger, without suggesting an identification. These works hint at a wealth of information still concealed in these spectra. 

Modelling of the X-shooter data has also revealed the spherical nature of the geometry of the early kilonova ejecta, contrary to expectations from most hydrodynamic models \citep{Sneppen2023}. Here, the fact that the $1\,\mu$m feature is a P~Cygni was a key discovery, as it allowed the geometry of the kilonova outflow to be assessed.%\LEt{Please check that I have retained your intended meaning.}
  
The identification of another P~Cygni feature is therefore important as it can be used to verify the \(1\,\mu\)m line and, because of the anticipated extreme wavelength-dependence of the opacity in \rproc-dominated ejecta, as a further probe of the geometry and abundance stratification of the kilonova. In this paper, we revisit these X-shooter spectra and identify a new spectral component, a P~Cygni feature at 760\,nm rest wavelength, which emerges most clearly several days post-merger.    %Due to the incompleteness of line-lists for \rproc elements, it is difficult unequivocally to assign a spectral origin. However, we investigate the most possible line identification. %, but its existence is significant to $x\sigma$ for both epochs 4 and 5. 
We investigate the wavelength stratification of the line and show that it provides an independent estimate of the expansion velocity of the ejecta, which is in good agreement with the constraints from the \(1\,\mu\)m feature. We also investigate whether this feature can be explained by the expected lines of yttrium at the temperature of the continuum fit and find that 4d\(^2\)--4d5p transitions of Y\(^+\) match the feature well. That is, in terms of central wavelength, velocity stratification, and the weaker prominence of the feature compared to its stronger Sr\(^+\) counterpart, this feature is well-described as a P~Cygni profile caused by transition lines from singly ionised yttrium. Furthermore, the fact that the P~Cygni feature emerges 3-4 days post-merger is a property of the yttrium identification, because at velocities of $0.2c-0.3c,$ the combined interference of different lines of Y\(^+\) conceals the characteristic P~Cygni spectral shape. 

\section{The 760\,n\titlelowercase{m} line detection }\label{sec:iden}
\begin{figure}
    \centering
    %Plot_spectra_and_fits-1.ipynb
    %\includegraphics[width=\linewidth,viewport=25 20 710 700 ,clip=]{Pictures_no_grid/750_spectrum_grid.pdf}
    \includegraphics[width=\linewidth,viewport=25 20 710 700 ,clip=]{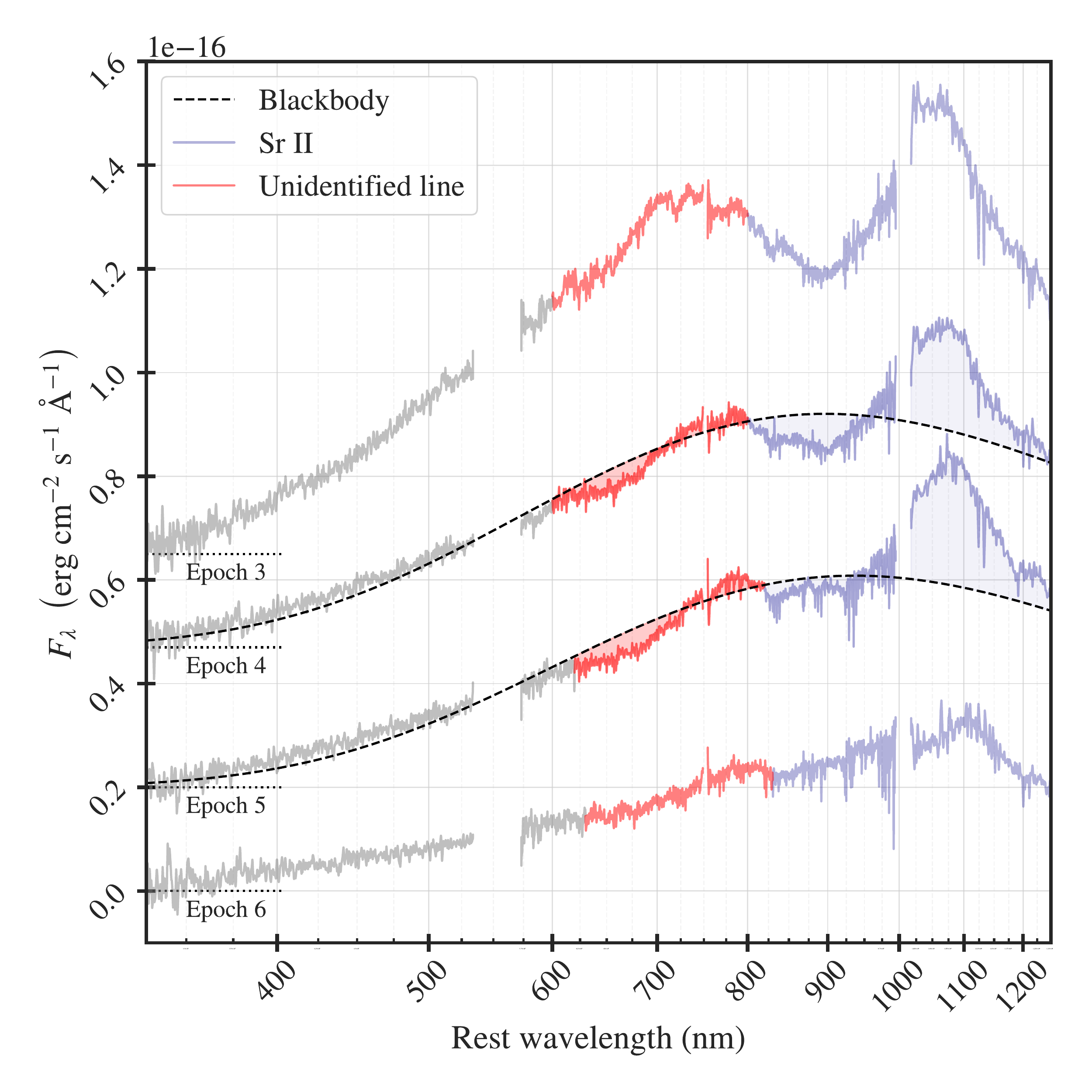}
    \caption{X-shooter spectra for epochs 3--6 (3.4-6.4 days post-merger) with blackbody fit and emphasis on the P~Cygni profiles of Sr\(^+\) (blue) and an unidentified spectral line at 760\,nm (red). For clarity, and to show them on the same plot, the spectra are offset as indicated by the dotted black lines. The feature is clearly noticeable from epoch 3 and onwards. However, the line is poorly constrained in epoch 3 as it less prominent relative to the continuum and overlaps with the most blueshifted parts of the Sr$^+$ P~Cygni. It is also poorly constrained in epoch 6 and later as the continuum is noisy and less well fit by a blackbody. Epochs 4 and 5 therefore provide the best constraints on the line properties.}
    \label{fig:750_spec}
\end{figure}

For this analysis, we use the X-shooter spectra, which were taken daily beginning 1.4 days after the merger, and follow the data reduction presented in \citet{Watson2019}. %observed and reduced as laid out in \citet{Watson2019} and \citet{Sneppen2023}. 
As illustrated in Fig.~\ref{fig:750_spec}, the optical spectra from epochs 3--6 (3.4--6.4 days post-merger) contain a clear excess of emission around 760\,nm with a corresponding absorption feature at about 650\,nm. The overall shape of the spectral component is reminiscent of P~Cygni features, which are characteristic of expanding envelopes where the same spectral line yields both an emission peak near the rest wavelength and a blueshifted absorption feature. While the feature is also there in epochs 3 and 6, we focus in this analysis on epochs 4 and 5, where the feature stands out most clearly and its properties can be most readily measured.% \citep{Jeffery1990}. 

%I open at the close. 

\begin{figure}
    \centering
    %Plot_spectra_and_fits-with-760Pcygni.ipynb
    %\includegraphics[width=\linewidth,viewport=15 15 465 565 ,clip=]{Pictures/750_spectrum_v_continuum_both_4.pdf}
    %\includegraphics[width=\linewidth,viewport=18 15 485 565 ,clip=]{Pictures/750_spectrum_v_continuum_both_4.pdf}
    \includegraphics[width=\linewidth,viewport=18 15 485 595,clip=]{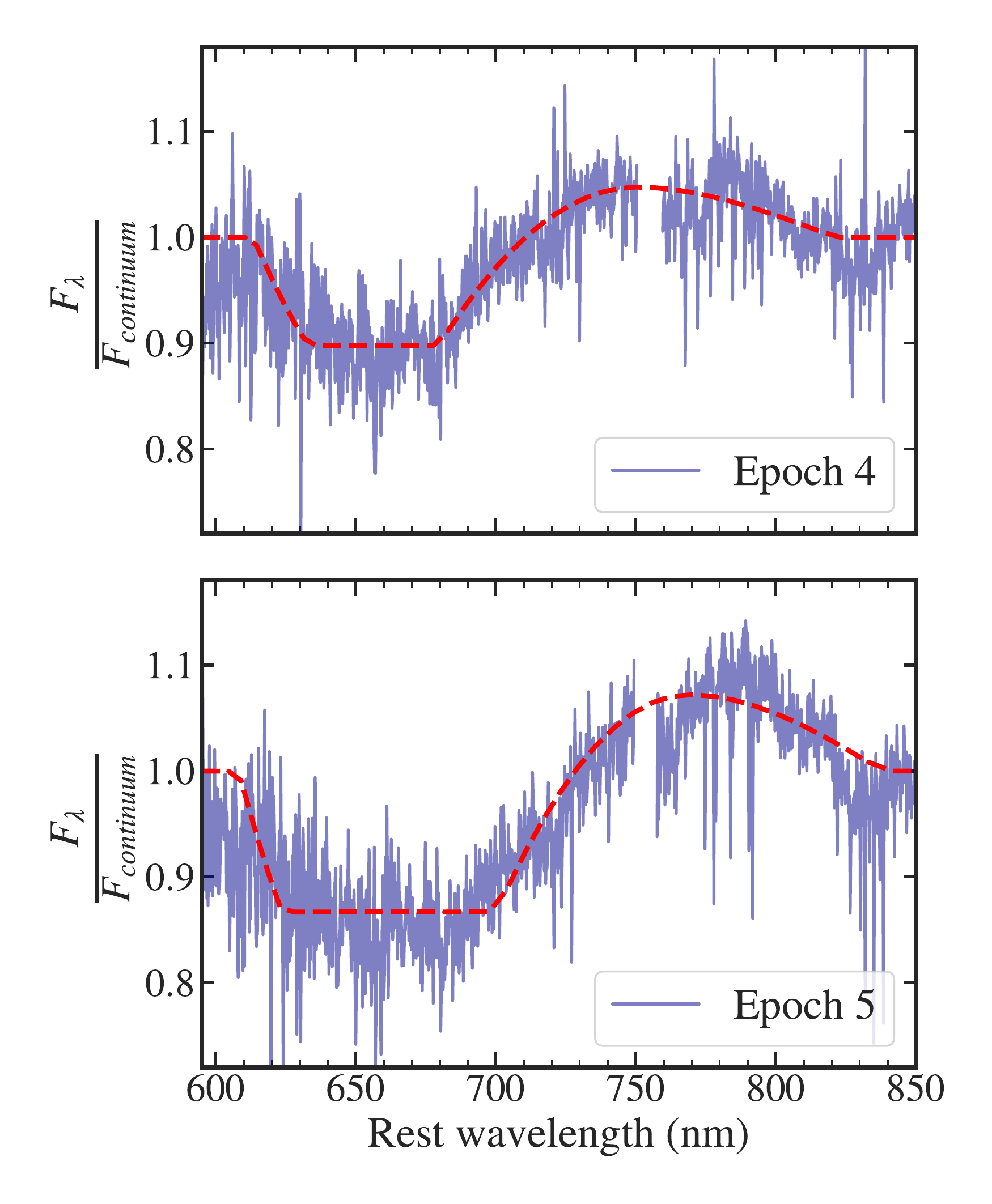}

    \caption{Spectra for epochs~4 and 5 divided by the model continuum flux, with single-line P~Cygni profiles fit around 760\,nm overlaid. For both epochs the existence of the feature is highly significant.}
    \label{fig:750_chi}
\end{figure}

%This plot does not contain lower limits
\begin{table*}
\renewcommand{\arraystretch}{1.5} 
\caption{Epochs~4 and 5 best-fit parameters for single-line P~Cygni (with free-to-vary central wavelength) and for the Yttrium P~Cygni model. This includes (1) the best-fit cross-sectional velocity inferred from the blackbody radii assuming Planck cosmology, $v_\perp$, and several parameters of the P~Cygni feature, including (2) the photospheric velocity, $v_{\rm phot}$, (3) the maximum outer ejecta velocity, $v_{\rm max}$, (4) the e-folding scale of the optical depth, $v_e$, (5) the optical depth of the line at the photospheric velocity, $\tau$, and (6) the central wavelength, $\lambda_0$. The uncertainties are derived from the 16th and 84th percentile and only include the statistical uncertainties in the fitting framework; i.e.\ they do not include the potentially significant systematic effects from line-blending and reverberation effects \citep[see][]{Sneppen2023_H0}. All velocities are given in units of the speed of light, \(c\). }\label{table:comparison}
\centering
\begin{tabular*}{0.99\textwidth}{@{}l @{\extracolsep{\fill}} cccccc@{}}
\hline \hline %& $T_{eff}$ & $v_{\rm ph}$ &
     & $v_{\rm \perp}$ & $v_{\rm phot}$ & $v_{\rm max}$ & $v_{\rm e}$ & $\tau$ & $\lambda_0$ [$\mu$m] \\ \hline 
%$v_\perp$                      &  $0.129\pm0.007$  & $0.109\pm0.006$  \\
%$v_\parallel(\rm Sr^+))$        &  $0.173\pm0.001$  & $0.146\pm0.001$ \\
\multicolumn{2}{l}{Single P~Cygni}   \\ 
Epoch 4 &  $0.129_{-0.007}^{+0.007}$ & $0.156_{-0.001}^{+0.001}$ & $0.183_{-0.001}^{+0.001}$ & $0.5_{-0.2}^{+0.3}$ & $0.54_{-0.02}^{+0.02}$ & $0.751_{-0.001}^{+0.001}$ \\
Epoch 5 &  $0.112_{-0.006}^{+0.006}$ & $0.191_{-0.004}^{+0.005}$ & $0.210_{-0.004}^{+0.005}$ & $0.7_{-0.2}^{+0.2}$ & $4.7_{-1.3}^{+3.1}$ & $0.769_{-0.001}^{+0.001}$ \\

%\multicolumn{2}{l}{A single P~Cygni fit} Hutsemekers  \\
%Epoch 4 &  $0.127\pm 0.007$ & $0.152\pm 0.001$ & $0.177\pm 0.001$ & $8.57\pm 1.4$ & $0.63 \pm 0.05$ & $0.736\pm 0.001$ \\
%Epoch 5 &  $0.112\pm 0.006$ & $0.178\pm 0.002$ & $0.198\pm 0.002$ & $9.8 \pm 7.9$ & $7.8\pm 6.5$ & $0.749\pm 0.001$ \\

%\multicolumn{2}{l}{760\,nm P~Cygni ()}   \\
\hline\hline
     & \(v_{\rm \perp}\) & \(v_{\rm phot}\) & \(v_{\rm max}\) & \(v_{\rm e}\) & \(\tau_{\rm avg}\) & \\ \hline 
\multicolumn{5}{l}{Y\(^+\) P~Cygni from 4d5p--4d\(^2\) line-transitions with \(\lambda \in [726\,{\rm nm}; 788\,{\rm nm}]\)}  \\
Epoch 4 & $0.129_{-0.007}^{+0.007}$ & $0.168_{-0.003}^{+0.004}$ & $0.220_{-0.002}^{+0.002}$ & $0.4_{-0.1}^{+0.1}$ & $0.161_{-0.011}^{+0.013}$ &  \\
Epoch 5 & $0.110_{-0.006}^{+0.006}$ & $0.176_{-0.002}^{+0.003}$ & $0.191_{-0.003}^{+0.003}$ & $0.6_{-0.2}^{+0.2}$ & $1.74_{-0.18}^{+0.21}$ &  \\
\hline
\end{tabular*}
\end{table*}

\subsection{Statistical significance}
To determine the significance of the feature, we compare the $\chi^2$ with and without fitting this spectral feature. For the null model, we fit the continuum as a blackbody as empirically proposed in many previous works \citep[e.g.][]{Malesani2017,Drout2017,Shappee2017,Waxman2018,Watson2019} and as theoretically and physically motivated in \cite{Sneppen2023_bb}. However, to account for the effects of spectral features on the continuum fit, we include parameters for the observed Gaussian emission lines at $1.5\mu$m and $2.0\mu$m and the Sr\(^+\) P~Cygni profile following the prescription in \citet{Watson2019}. The analysis is robust regardless of whether these spectral features are parameterised in the fit as we do here or the analysis is limited to wavelengths without these spectral components \citep{Sneppen2023}. To complete the full model, in addition to continuum and nuisance spectral components, we add a P~Cygni prescription to model the 760\,nm line. 

For this analysis, we use the implementation of the P~Cygni profile in the Elementary Supernova Model\footnote{The adopted implementation is Ulrich Noebauer's \texttt{pcygni\_profile.py} \url{https://github.com/unoebauer/public-astro-tools}.}, where the profile is set by several properties of the kilonova atmosphere. The profile is expressed in terms of the rest wavelength, $\lambda_0$, and the line optical depth, $\tau$, with a velocity stratification parameterised by a scaling velocity, $v_e$, a photospheric velocity, $v_{\rm phot}$, and a maximum ejecta velocity, $v_{\rm max}$. This parameterisation assumes an optical depth with an exponential decay in velocity as presented in \cite{Thomas2011}, but one can equivalently assume a power-law decay with the resulting constraints being indistinguishable. In short, the optical depth determines the strength of absorption and emission, while the velocity of the ejecta sets the wavelength stratification.

For epochs~4 and 5, no blackbody continuum model can fit the observed profile well from 600 to 800\,nm (as seen in Fig \ref{fig:750_chi}), with the reduced chi-squared of $\chi_{\rm \nu,4th}^2=1.65$ and $\chi_{\rm \nu,5th}^2=1.88$. In contrast, fitting the residual structure with the additional P~Cygni model instead yields $\chi_{\rm \nu,4th}^2=1.17$ and $\chi_{\rm \nu,5th}^2=1.21$ over the same wavelengths (see the posterior landscape in Appendix Fig. \ref{fig:4posterior} and \ref{fig:5posterior}). The fact that $\chi_\nu^2>1$ for the full model may indicate that the errors are underestimated, that a single P~Cygni prescription is an over-simplification, or that there are unknown spectral substructures. 

Nevertheless, the substantial improvement in $\chi^2$ while only requiring the five additional parameters of the P~Cygni suggests that the feature is highly significant and we investigate this further here. 
%is justified by the Deviance Information Criterion \citep[see e.g.][]{Liddle2007}. 
To examine the significance using more observationally related uncertainties, we artificially inflate the errors of the full model so that $\chi^2_\nu=1$. In this case, the simpler continuum model without the 760\,nm P~Cygni provides a poorer fit to the data, with $\chi^2$ values going from 14\,321 to 10\,102 and 15\,686 to 10\,096 for epochs 4 and 5, respectively. This results in $\Delta\chi^2=4219$ and 5590 for five additional degrees of freedom. This corresponds to a null-hypothesis probability rejected at 26$\sigma$ and 33$\sigma$, again for epochs 4 and 5, respectively. One could also apply the deviance information criterion \citep[DIC; see e.g.][]{Liddle2007} here, which results in a \(\Delta\mathrm{DIC} < -4000\) for both epochs corresponding to decisive evidence on the Jeffreys scale \citep[e.g.][]{Jeffreys1939,Kass1995}.
%4215 & 4708

\begin{figure*}[h]%[ht!]
    \centering
    \includegraphics[width=0.48\linewidth]{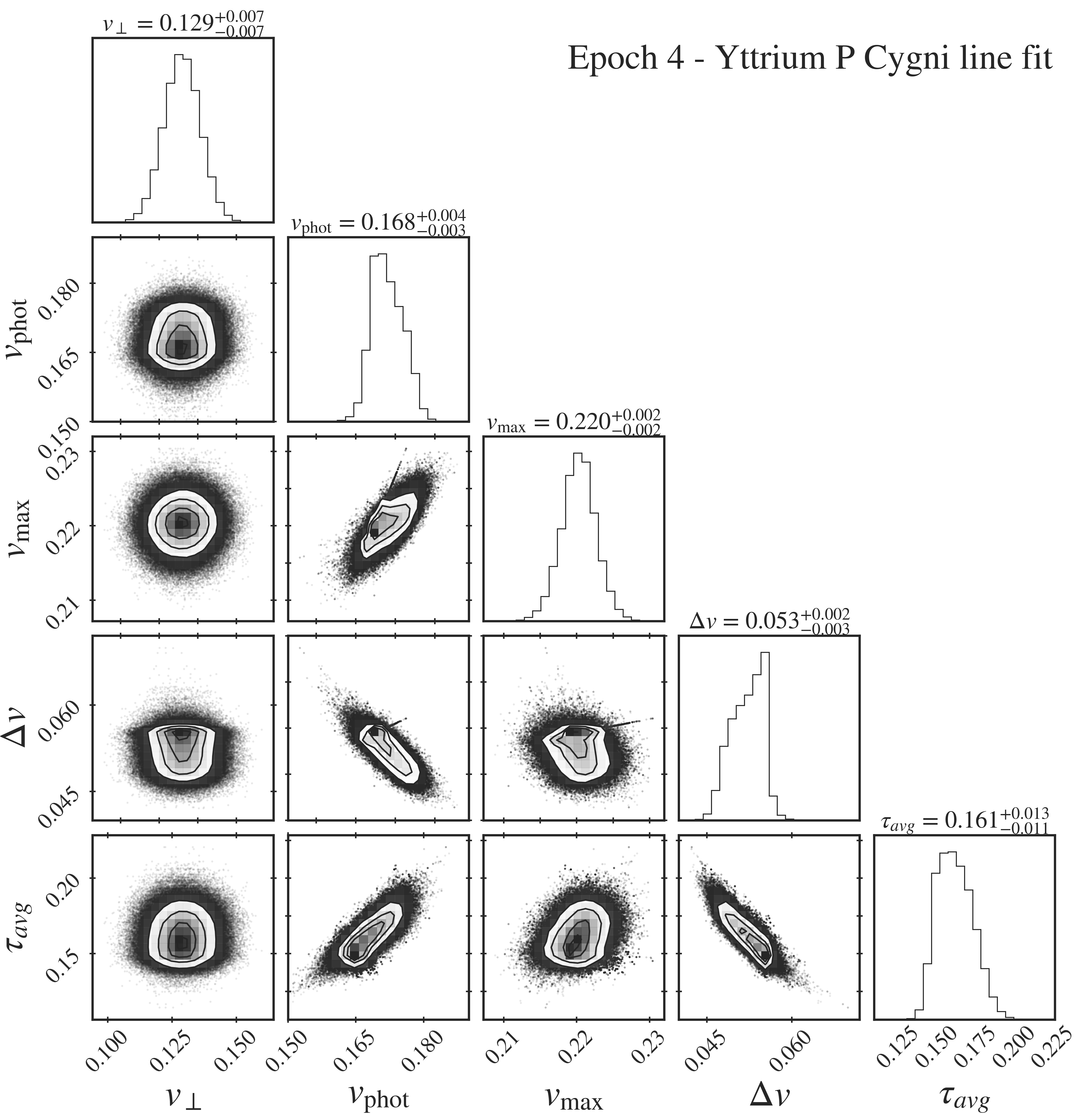}
    \includegraphics[width=0.48\linewidth]{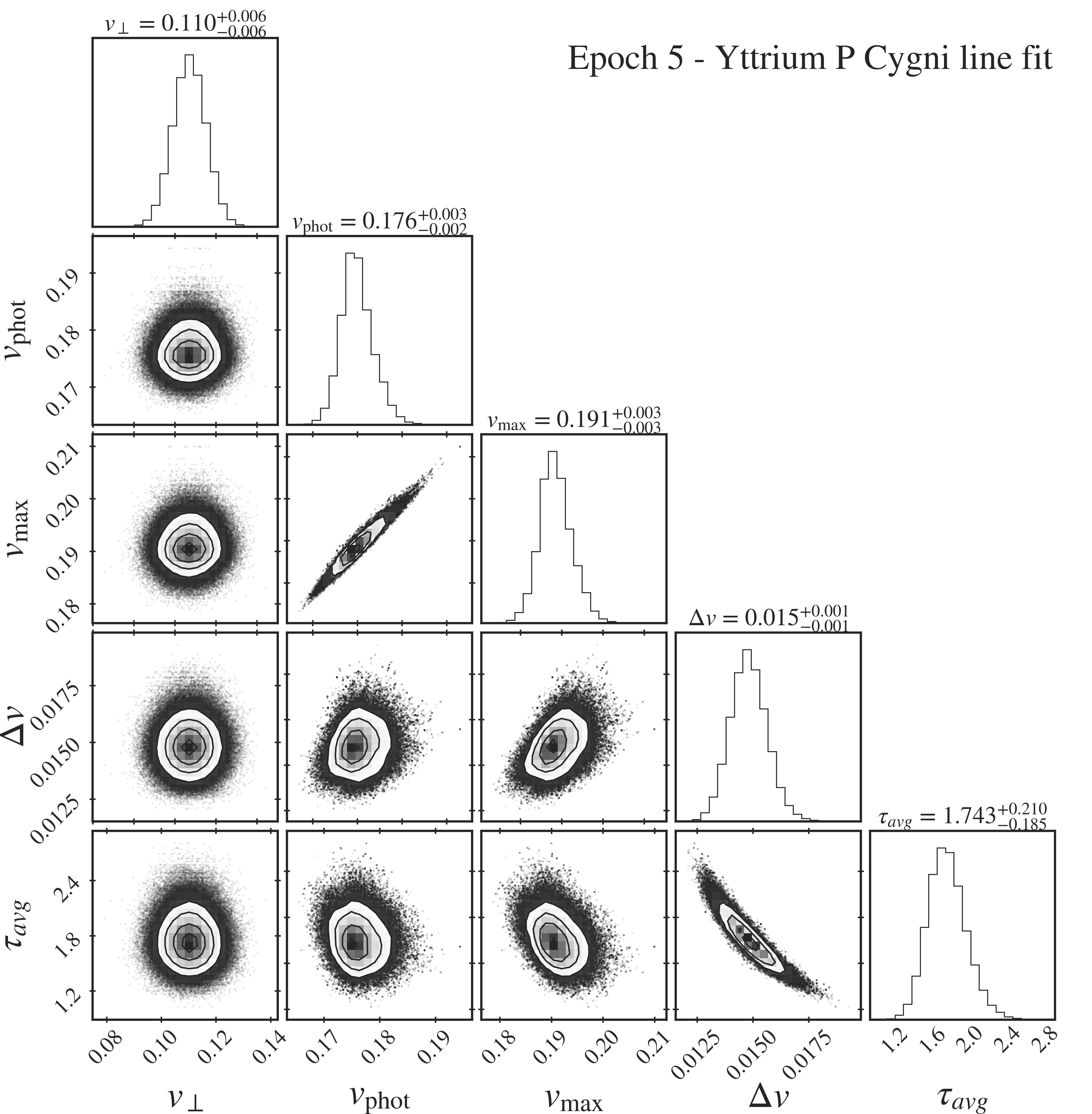}
    
    \caption{Corner plots showing the posterior probability distribution of key parameters for epochs~4 (\emph{left}) and 5 (\emph{right}) fitting as a P~Cygni feature from the set of Y\(^+\) lines from the $4{\rm d}5{\rm p}$-$4{\rm d}^2$ transitions with $\lambda \geq 700$\,nm. %The posterior distributions have a single global minimum for epoch~4, while the fitted line is more saturated in epoch~5, so the constraining power is much lower. 
    The fit parameters are (1) the cross-sectional velocity, $v_\perp$, derived from the blackbody normalisation fit assuming Planck cosmology \citep{Planck2018}, (2) the photospheric velocity, $v_{\rm phot}$, (3) the maximum outer ejecta velocity, $v_{\rm max}$, (4) the velocity range, $\Delta v = v_{\rm max}-v_{\rm phot}$, and (5) the average optical depth of the Y\(^+\) lines weighted by their relative strength, $\tau_{avg}$. Velocities are given in units of the speed of light, \(c\). The central wavelength of a line, $\lambda_0$, and its relative strength is set by the $4$d$5$p to $4$d$^2$ transitions in Y\(^+\) (see \cite{Biemont2011} and Table \ref{table:Yttrium}) in LTE at about 0.4\,eV. These constraints show only the statistical uncertainty in the fitting model.}
    \label{fig:yttriumposterior}
\end{figure*}

%$7780_{-160}^{+60}$
\subsection{The central wavelength}
The best-fit central wavelength of the single P~Cygni line fit is $750.5 \pm 1$\,nm and $769.1 \pm 1.2$\,nm for epochs~4 and 5, respectively. The unweighted average of these would indicate a central wavelength of roughly $760 \pm 10$\,nm. While the statistical uncertainties of fitted parameters from each epochs are very small and is even at a level where the central wavelength value is inconsistent between the two epochs, we emphasise that there are systematic effects that are expected to bias the central value significantly in one or both of the epochs. 
First, without a prior spectral identification, we assume the origin of the line to be a single transition, when a blending of nearby lines may produce a similar profile---indeed, we think multiple transitions from related energy levels are highly likely (see Sect.~\ref{sec:line-id}). 
Second, the most redshifted part of the 760\,nm feature is coincident in wavelength with the most blueshifted absorption from Sr\(^+\), and so the exact location of the emission peak and the maximum ejecta velocity of the Sr line-forming region are degenerate. 
Third, the relativistic velocity of the ejecta moving at $\approx 0.15c$ suggest that there are significant time delays, which implies the different wavelengths of the profile may be probing different times, that is,\ at different temperatures and optical depths.

\subsection{Velocity stratification}\label{sec:vel}
In addition to the central wavelength, the P~Cygni profile also provides information on the properties of the expanding ejecta. The spherical nature of the early ejecta of AT2017gfo was first shown by \citet{Sneppen2023} from the high degree of consistency between the line-of-sight velocity, $v_\parallel$, from the Sr\(^+\) lines and the cross-sectional velocity, $v_\perp$, derived from the continuum flux. Here, the cross-sectional velocity (or equivalently area) is also determined by comparing the observed flux with the intrinsic blackbody luminosity, while assuming Planck cosmology \citep{Planck2018} and the Hubble flow recession velocity of the host galaxy in \citet{Mukherjee2021}. In addition, the line fits for $\textrm{Sr}^+$ suggest the photospheric velocity is $v_\parallel(\textrm{Sr}^+) \approx 0.17c$ and $0.15c$ in epochs~4 and 5, respectively \citep{Sneppen2023}. This allows the degree of sphericity to be parameterised in terms of the asymmetry index: 
\begin{equation}
    \Upsilon( \textrm{Sr}^+) = \frac{v_\perp-v_\parallel(\textrm{Sr}^+)}{v_\perp+v_\parallel(\textrm{Sr}^+)}
.\end{equation}

Similarly, the new line profile also contains spectral constraints on the velocity stratification of the expanding ejecta, which is shown in the posterior probability distributions of Fig.~\ref{fig:yttriumposterior}.
As seen in Table~\ref{table:comparison}, the photospheric velocity derived from the line, $v_{\parallel}{\rm(760\,nm)} \approx 0.16c$, is larger than the cross-sectional velocity of the continuum and broadly similar to that derived from the Sr\(^+\) lines. %with $\Upsilon_{\rm 4th}(0.76\mu \textrm{m}) = -0.08 \pm 0.02$ and $\Upsilon_{\rm 5th}(0.76\mu \textrm{m}) = -0.17 \pm 0.04$. 
Therefore, this new line seems to confirm an increasing discrepancy in later epochs between the velocity indicators with $v_\perp<v_\parallel$ as was hinted at by the Sr\(^+\) P~Cygni analysis \citep{Sneppen2023}. However, this does not necessarily imply an increasingly prolate geometry; for example, the blackbody assumption is a worsening approximation in later epochs, resulting in a substantial increase in systematic uncertainty.  % velocity-stratification from Strontium. 

We note that the fits to the line for epoch~5 appear saturated. However, there is a strong degeneracy between the optical depth and the thickness of the line-forming region (e.g.\ Fig.~\ref{fig:yttriumposterior}). Therefore, the apparent saturation could be caused by a thin line-forming region in the fit (i.e.\ a small difference between $v_{\rm phot}$ and $v_{\rm max}$) producing a weak feature, which in the best fit is compensated by increasing the optical depth. However, it is likely that line blending, reverberation effects, or other systematic effects may change the velocity stratification and, by extension, the optical depth. Additionally, we note a high-opacity line is problematic for any direct comparisons between the photospheric velocity and the cross-sectional velocity, as a high opacity may detach the line-forming region from the continuum, inducing a bias in the inferred photospheric velocity \citep{Sim2017}.

\section{Yttrium identification}
\label{sec:line-id}
The new P~Cygni line detection is useful, not only because it allows comparison with the velocity analysis of Sr\(^+\), but also because it provides a tantalising potential identification, which may subsequently improve the constraint on the composition and \rproc synthesis of kilonovae. However, any identification is difficult, as current atomic line lists are incomplete and/or suffer from significant inaccuracies in wavelength and/or line strength \citep[e.g. VALD, Kurucz;][respectively]{Ryabchikova2015,Kurucz2018}, especially for elements beyond the first \rproc peak. Due to the difficulty in generating this information for heavier elements, despite significant recent efforts \citep[e.g.][]{Tanaka2020}, line lists with reliable wavelengths and transition strengths remain mostly incomplete beyond the first peak, and especially longward of $1\,\mu$m, which hampers modelling efforts. 

We examined the line lists of the most obvious spectral candidates, that is, those\ with strong transitions due to a small number of valence electrons and low-lying energy levels \citep{Watson2019,Domoto2022}. First, the elements near the first \rproc peak, which are abundant and contain the only known prior identification, namely of Sr. Of these elements, neither Ru, Sr, or Zr have strong lines at the relevant wavelengths for their likely dominant ionisation states. There are several noteworthy strong lines from \rproc second peak elements (notably Ba\(^+\) and La\(^+\)), but these are located at considerably shorter (or longer) wavelengths.

\begin{figure}
    \centering
    %Plot_spectra_and_fits-with-760Pcygni.ipynb
    %\includegraphics[width=\linewidth,viewport=15 15 465 565 ,clip=]{Pictures/750_spectrum_v_continuum_both_4.pdf}
    \includegraphics[width=\linewidth,viewport=18 15 625 505 ,clip=]{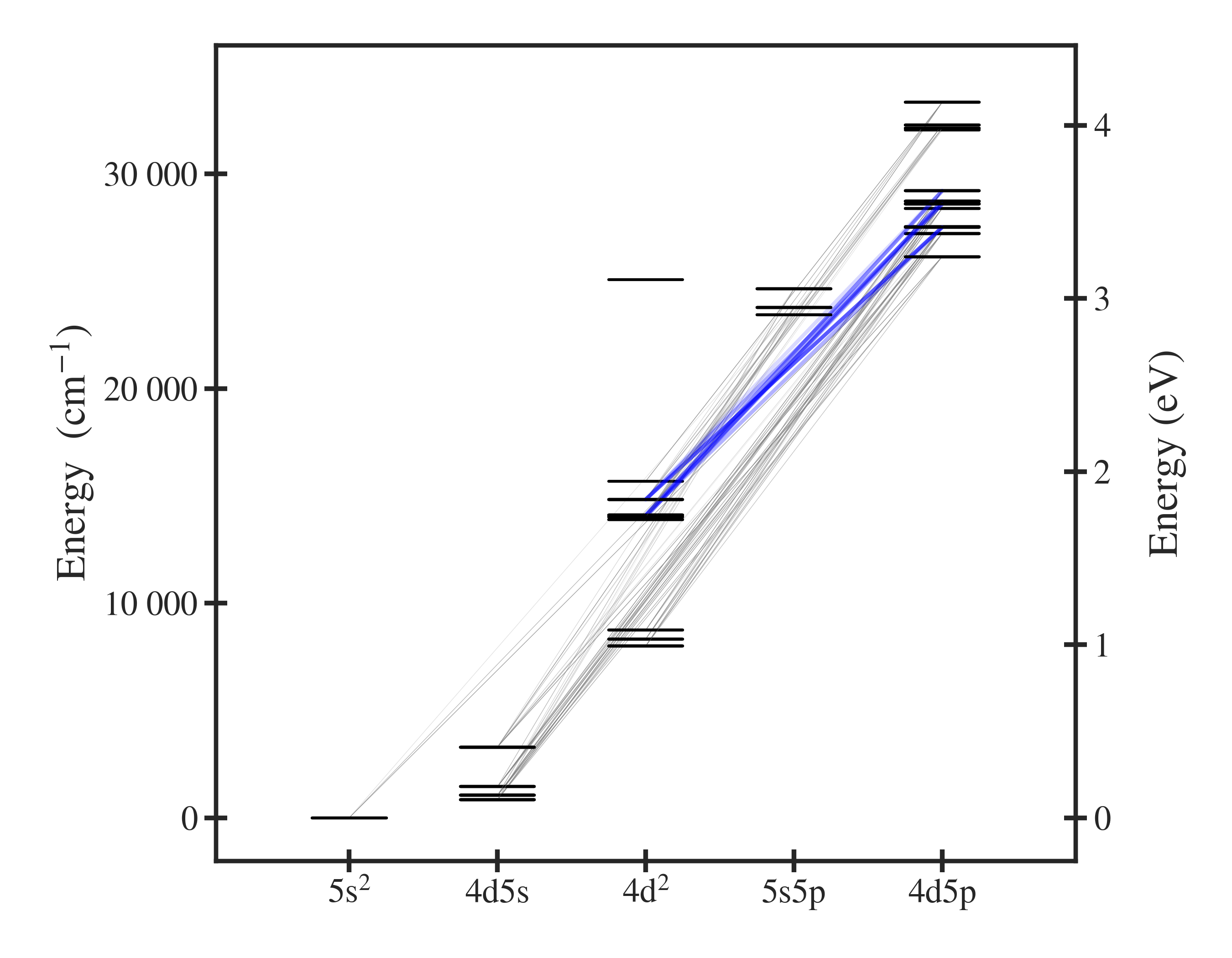}
    \caption{ Grotrian diagram showing the lowest energy levels in Y\(^+\). The transitions between energy levels are indicated, with the line transparency set by the transition strength, $\log{gf}$. The 4d5p--4d$^2$ (28\,000\,cm\(^{-1}\)--14000\,cm\(^{-1}\)) transitions are shown with thick lines in blue. Radiative transfer modelling analysis in \citep{Vieira2023} tentatively links the 4d5p--4d$^2$ (28\,000\,cm\(^{-1}\)--8000\,cm\(^{-1}\)) transitions to absorption at $\lambda < 450$\,nm in the epoch~1 spectrum; other notable transitions are around 300\,nm, too short to be observed by X-shooter (4d5p--4d5s, 28\,000\,cm\(^{-1}\)--1000\,cm\(^{-1}\)), and signatures of the 5s5p--4d$^2$ transitions (24000\,cm\(^{-1}\)--14000\,cm\(^{-1}\)) at 1$\mu$m, which are much weaker than the Sr\(^+\) line transitions at similar wavelengths. The 4d5p--4d$^2$ transitions have lower oscillator strength than several of the Y\(^+\) lines at shorter wavelengths, but due to the rapid decline of opacity with wavelength for typical \rproc elements, these transitions have higher opacity compared to the continuum at 760\,nm. When no lines are drawn between levels, the transition is negligible with $\log{gf} < -2$.
    }
    \label{fig:grotrian}
\end{figure}

\begin{table}
\caption{4d5p to 4d\(^2\) transitions for Y\(^+\). All transitions are used for the \textsc{tardis} modelling, while we tested the Y\(^+\) P\,Cygni framework using only the transitions listed here with $\lambda>700$\,nm and all the transitions in this table. Vacuum wavelengths ($\lambda$), energy levels ($E$), and oscillator strengths ($\log{gf}$) for the Y\(^+\) 4d5p to 4d\(^2\) lines are from \citet{Biemont2011}.}\label{table:Yttrium}
\centering
\renewcommand{\arraystretch}{1.25} 
\begin{tabular}{@{}c|cc|cc|c@{}} \hline\hline %& $T_{eff}$ & $v_{\rm ph}$ &
$\lambda$ & \multicolumn{2}{|c|}{Lower level} & \multicolumn{2}{|c|}{Upper level} & $\log{gf}$ \\
 (nm) & $4{\rm d}^2$ & $E$ (cm$^{-1}$) & $4{\rm d}5{\rm p}$ & $E$ (cm$^{-1}$) & \\ \hline 
788.19  & $b^1 D_2$  & 14\,832  & $z^1 P_1^\circ$ & 27\,517 & $-0.60$ \\
745.03  & $a^3 P_2$  & 14\,098  & $z^1 P_1^\circ$ & 27\,517 & $-1.44$ \\
733.29  & $a^3 P_0$  & 13\,883  & $z^1 P_1^\circ$ & 27\,517 & $-1.98$ \\
726.42  & $b^1 D_2$  & 14\,832  & $z^3 D_1^\circ$ & 28\,595 & $-1.11$ \\[3pt]
\hdashline
689.60  & $a^3 P_2$  & 14\,098  & $z^3 D_1^\circ$ & 28\,595 & $-1.67$ \\
685.82  & $a^3 P_1$  & 14\,018  & $z^3 D_1^\circ$ & 28\,595 & $-1.69$ \\
683.24  & $a^3 P_2$  & 14\,098  & $z^3 D_2^\circ$ & 28\,730 & $-1.86$ \\ %
679.54  & $a^3 P_0$  & 13\,883  & $z^3 D_1^\circ$ & 28\,595 & $-1.54$ \\
679.53  & $a^3 P_1$  & 14\,018  & $z^3 D_2^\circ$ & 28\,730 & $-1.03$ \\ %-
661.37  & $a^3 P_2$  & 14\,098  & $z^3 D_3^\circ$ & 29\,213 & $-0.83$ \\
\hline
\end{tabular}
\end{table}

The most promising candidate, yttrium, should ---at these temperatures and densities--- principally be in the singly ionised state, Y\(^+\), which has strong line transitions from $4$d$5$p to $4$d$^2$. These transitions have an LTE-weighted mean wavelength of around 760--770\,nm with the most prominent line being at 788.19\,nm (see Table~\ref{table:Yttrium} and Fig.~\ref{fig:grotrian}). Given the strong transition, its relatively high expected abundance, and its atomic number proximity to Sr, yttrium seems a likely candidate. Interestingly, a specific prediction of the yttrium identification is that the feature should first emerge clearly 3-4 days post merger, as is indeed observed in the spectra. This is because in earlier epochs with greater characteristic velocities, the absorption from the lines around 760\,nm will coincide in wavelength with the emission of the 670\,nm Y\(^+\) lines, and act to suppress the prominence of the feature, as discussed further in Sect.~\ref{sec:discussion}.

The ionisation energies for Y and Y\(^+\) are 6.2 and 12.2\,eV, respectively. Using the Saha equation and the range of electron densities expected for the ejecta ($n_e \approx 10^{6}-10^{8}\,\rm{cm}^3$) suggests characteristic temperatures for the first and second ionisation of 2200--2400\,K and 4100--4700\,K, respectively. For the observed temperature in the analysed epochs, $T \approx 3000$\,K, the fraction of yttrium in the singly ionised state should be near unity. We note that for earlier epochs (for instance 1.4 days post-merger), the inferred temperature is near the second ionisation threshold, which suggests a substantial fraction of yttrium at those epochs may be present as Y\(^{++}\). The growing prominence of this feature over the epochs could in part be the result of the decreasing ionisation of yttrium in LTE. However, given the presence of a strong Sr\(^+\) absorption in the first epoch spectrum \citep{Watson2019}, and the lower first and second ionisation energies of Sr, decreasing ionisation alone is not a very satisfactory explanation for the increased prominence of the 760\,nm feature at epochs~3--6.

\begin{figure*}[h]%[ht!]
    \centering

    \includegraphics[width=0.49\linewidth, viewport=15 5 525 505, clip=]{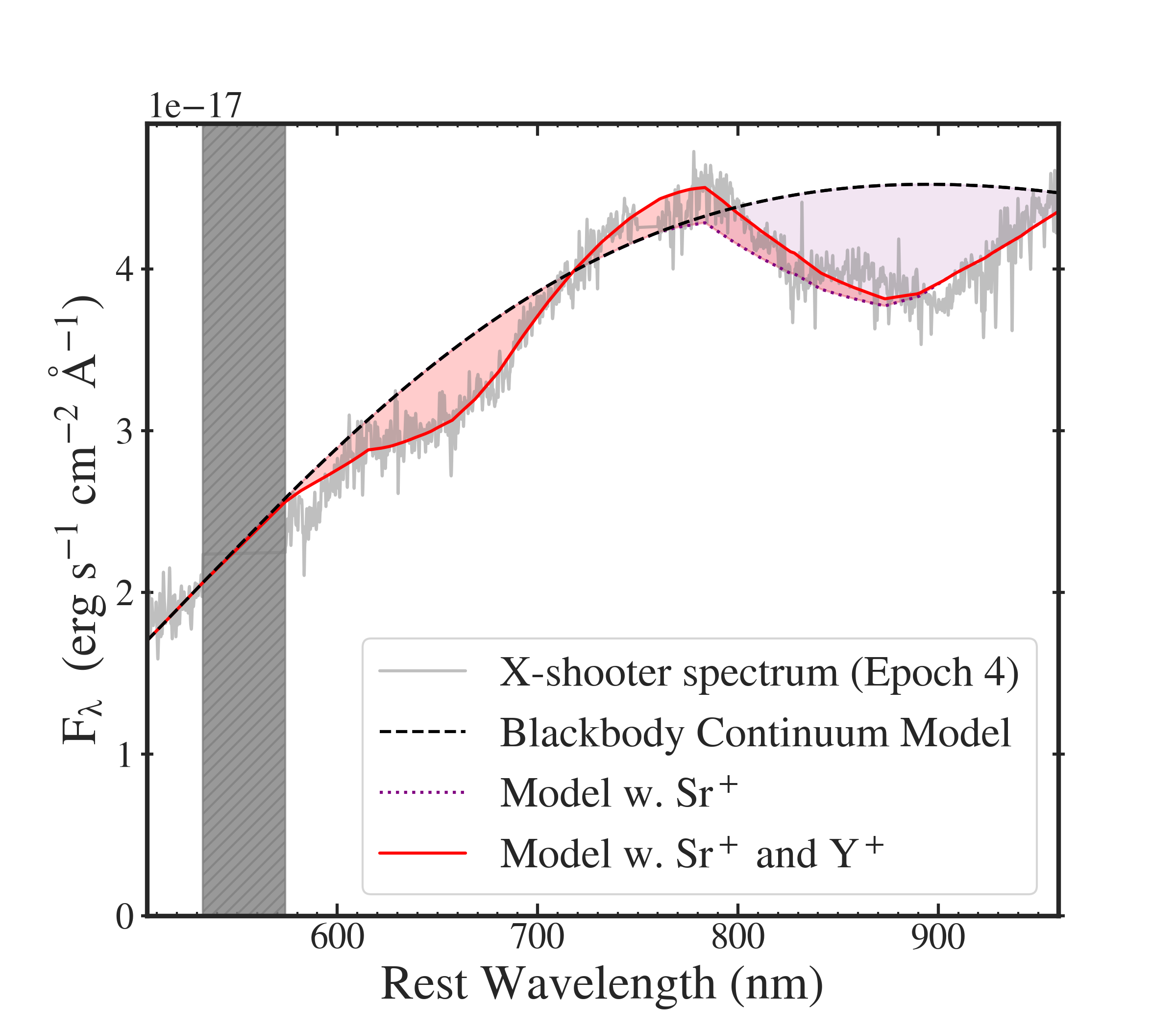}
    \includegraphics[width=0.49\linewidth, viewport=15 5 525 505, clip=]{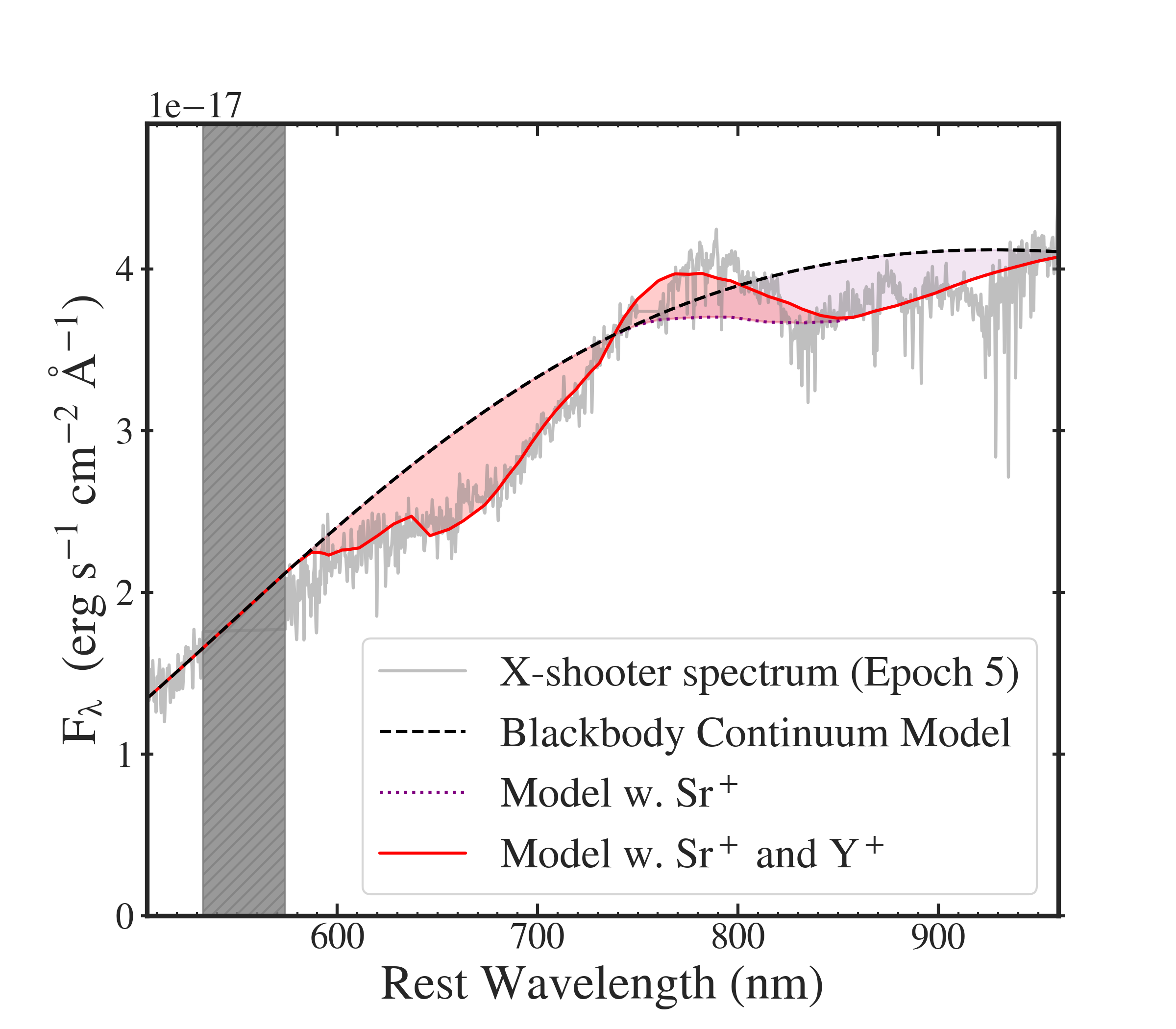} %, viewport=15 15 705 725, clip=
    \caption{X-shooter spectra with model fit including Y\(^+\) lines with $\lambda \geq 700$\,nm for epochs 4 and 5. The blackbody continuum model is shown as a dashed black line, the blackbody plus the Sr\(^+\) P~Cygni lines as a dotted purple line, and the full model including both Sr\(^+\) and Y\(^+\) as a solid red line. The partially transparent purple and red regions indicate the successive contributions to these models of the Sr\(^+\) and Y\(^+\) P~Cygni lines, respectively. The grey hatched region indicates the overlapping noisy regions between the UVB and VIS arms of the spectrograph. The Y\(^+\) lines (from the 4d\(^2\)--4d5p transitions) have fixed central wavelengths and their relative strengths are set by the LTE condition.}
    \label{fig:Yttrium}
\end{figure*}

\subsection{P~Cygni analysis of the yttrium feature}

In Fig.~\ref{fig:Yttrium}, we show the continuum and P~Cygni fit for the Y\(^+\) line transitions with $\lambda \geq 700$\,nm. We note that these transitions provide a good fit to the observed absorption trough and emission peak. This is despite the fact that, for this fit, we fixed the central wavelength and the relative strengths of the lines according to an LTE calculation based on the atomic line data of \citet{Biemont2011}. The multi-line nature of the fit even allows the spectral substructures to be fit, that is, tracing both the overall line shape but also the small-scale wiggles observed, for example, in the epoch 5 fit. Interestingly, this shows that characteristic wiggles may be produced by the combination of nearby lines with varying transition strengths, which in turn suggests that spectral fingerprints of specific \rproc elements could be hiding within other such wiggles of the data. We note that including shorter wavelength transitions generally leads to smaller fitted velocities (as these transitions naturally extend the profile to slightly shorter wavelengths), but the exact lower wavelength cutoff below which lines are not considered in the fit is unimportant in terms of the goodness-of-fit. Without substantially changing the best-fit profile-shape, we can include all $4{\rm d}5{\rm p}$-$4{\rm d}^2$ transitions, but as the $\sim 670$\,nm lines are not the driving lines behind the observed feature and there are strong indications that these shorter wavelength features may be suppressed by line blanketing (see Sec. \ref{sec:tardis}), for the presented fit we emphasise the dominant higher-wavelength transitions. In Sect. \ref{sec:tardis}, we use the full line list and explore the relation between the $\sim670$\,nm and $\sim760$\,nm transitions. 

The column density inferred from fitting the P~Cygni Y\(^+\) 4d5p-4d\(^2\) transition fits suggests total yttrium masses of $0.6\times10^{-4}\,\rm{M}_{\odot}$ and $2.4\times10^{-4}\,\rm{M}_{\odot}$ would be required to produce these features for epochs 4 and 5, respectively, assuming spherical symmetry. The derived masses should be treated with caution as there is no correction for light-travel-time effects or additional blending lines, which could also potentially bias inferred properties. These are lower limits on the total amount of material ejected, as they trace only the matter between the photospheric front and the outer atmosphere. This is especially apparent for the best-fit in epoch 5, where the line profile is saturated and the line-forming region may therefore be detached from the inner emitting mass. Given the total mass within the atmosphere ejected inferred from light-curve modelling, $M_{\rm ej} \approx 3-5\times10^{-2}\,\rm{M}_{\odot}$ \citep{Kasen2017}, this would suggest a lower bound on the mass fraction of yttrium of around 0.5--0.8\% from epoch 5. This is within the yield estimated from numerical simulations of neutron-star mergers \citep{Perego2021}, and is consistent with estimates of the \rproc fraction in the Sun \citep{Bisterzo2014,Lodders2009}.    

\begin{figure*}
    \centering
    %Zirconium/Data
    %\includegraphics[width=\linewidth,viewport=18 22 490 520 ,clip=]{Pictures/Yttrium_tardis_2.png}
    %\includegraphics[width=\linewidth,viewport=18 22 1000 520 ,clip=]{Pictures/Yttrium_tardis_2_panel_n.png}
    \includegraphics[width=\linewidth,viewport=18 22 980 520 ,clip=]{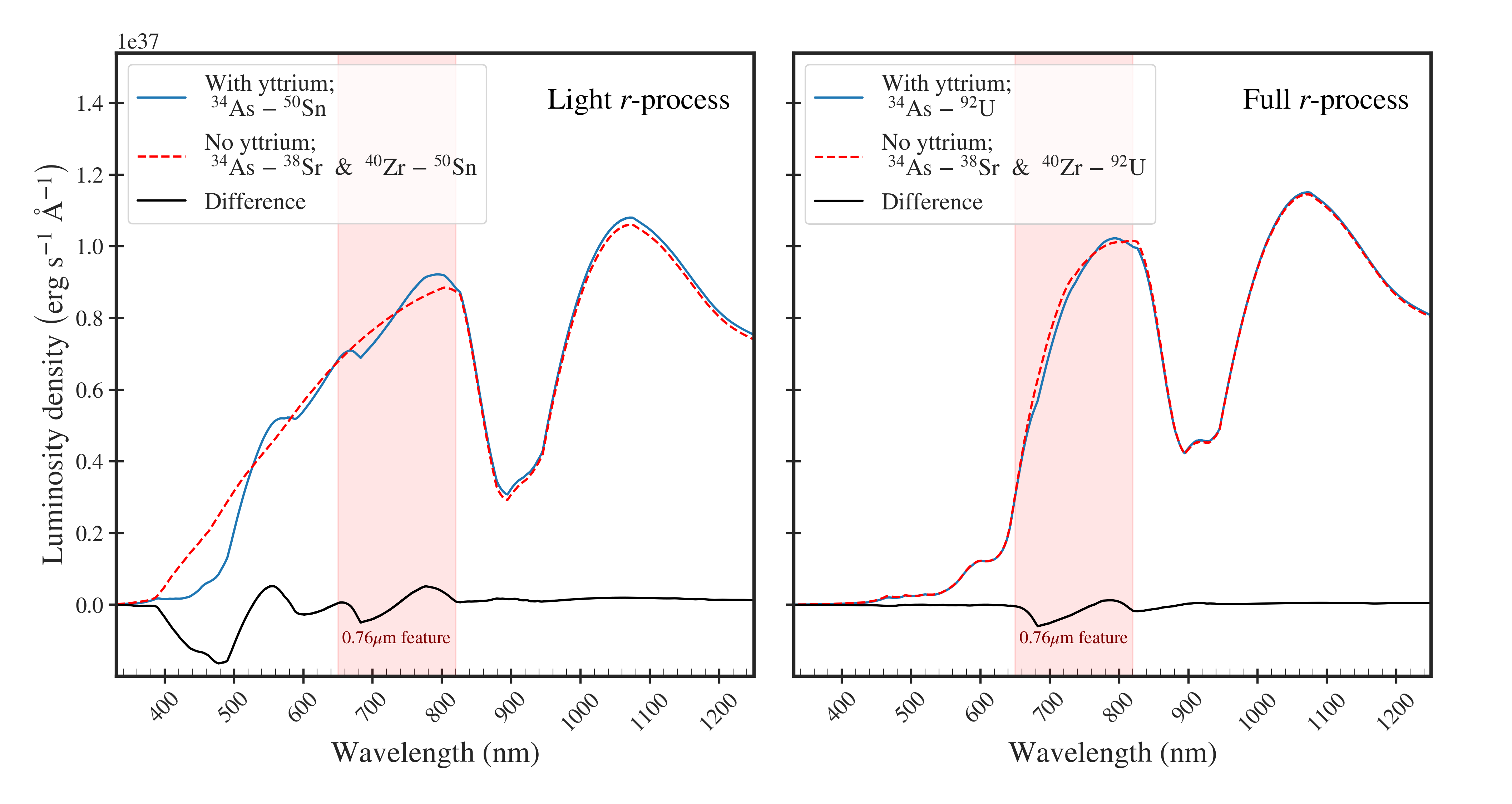}
    \caption{ Example \textsc{tardis} model spectra including yttrium (blue line) and excluding yttrium (red dashed line) but otherwise composed of \rproc elements with solar \rproc abundance ratios \citep{Lodders2009,Bisterzo2014}. The left panel is limited to the first \rproc peak elements ($34 \leq Z \leq 50$), while the right panel shows a model with all \rproc elements ($33 \leq Z \leq 92$).
    The spectrum at the photosphere is modelled as a blackbody, which is then propagated through a relativistically expanding atmosphere ($0.15c \leq v \leq 0.2c$) using the radiative transfer equations with a radially uniform composition using the \citet{Kurucz2018} line list. The red region indicates the 760\,nm feature, which is clearly associated with the presence of yttrium. The yttrium abundance in the model constitutes 0.8$\%$ of the total mass. The 0.8--1$\mu$m feature is the Sr\(^+\) P~Cygni feature \citep{Watson2019}. }
    \label{fig:tardis_750_spec}
\end{figure*}

\subsection{\textsc{tardis} analysis of the yttrium feature}\label{sec:tardis}

In addition to the P~Cygni fitting framework, we also test the prominence of features due to the proposed Y$^+$ transitions using the Monte Carlo radiative-transfer spectral synthesis code \textsc{tardis} \citep{Kerzendorf2014}. In Fig.~\ref{fig:tardis_750_spec}, we show the spectra resulting from a \textsc{tardis} simulation with a solar \rproc element abundance distribution for the \rproc elements below the second peak ($34 \leq Z \leq 50$). Our solar \rproc distribution is based on the solar abundance \citep{Lodders2009} with the $s$-process abundance distribution subtracted \citep{Bisterzo2014}. In the same figure, we also show the spectrum resulting from a simulation with the same parameters but without yttrium, as well as the difference between the models with and without yttrium. When yttrium is included, a P~Cygni component is visible at 760\,nm. A similar plot, but this time including all \rproc elements ($34 \leq Z \leq 92$), is also shown in Fig.~\ref{fig:tardis_750_spec}. In both scenarios, that is, with all elements or with only the lighter ones, the P~Cygni feature at about 760\,nm is still produced by the 4d$^2$--4d5p transitions of Y$^+$.

In addition to the $\sim760$\,nm feature that we observe, the presence of yttrium also results in (1) strong absorption of flux at short wavelengths $\leq 450$\,nm, (2) a small P~Cygni feature at about 650\,nm in emission (the presence or absence of which is hard to constrain observationally from the X-shooter spectra as the absorption sits in the noisy region between UVB and VIS arms of the spectrograph), and (3) a near negligible contribution to the 1$\mu$m feature from the $5$s$5$p-$4$d$^2$ transitions. The relative strengths of these features are affected by the opacity from other lines in the same region of the spectrum. This is seen most clearly in the difference between the left and right panels of Fig.~\ref{fig:tardis_750_spec}, where the more blueward features are significantly less prominent with the higher blue-wavelength opacity introduced to the radiative transfer model by the \rproc second peak elements. The relative prominence of the $\sim760$\,nm feature compared to its lower-wavelength counterparts could therefore be considered tentative evidence of a significantly higher opacity below $\sim700$\,nm, which is consistent with the arguments in \citet{Gillanders2022} that for synthetic models to match the observed flux distribution for $\lambda<750$\,nm in later epochs requires the inclusion of lanthanides. However, as we cannot confidently exclude a feature at $\lambda \sim600$\,nm, and because the line lists we currently have are incomplete --- a more complete line list could well result in greater blue-wavelength opacity from the lighter elements (with $34 \leq Z \leq 50$) --- we do not regard the dominance of the 760\,nm feature as definitive evidence of lanthanides; nevertheless, in agreement with \citet{Gillanders2022}, we do find it suggestive. Regardless, it is interesting to note the unique prominence of the P~Cygni feature from Y$^+$ in these models as well as the relative strength compared with the Sr$^+$ feature. We emphasise that the potential identification of a 760\,nm P~Cygni feature from Y\(^+\) noted here is complementary to, and independent of, a deficiency of flux noted at short wavelengths in early epochs, which has been tentatively suggested to be due to Y\(^+\) and, to a lesser degree, to Zr\(^+\) \citep{Vieira2023}. We note that \citet{Vieira2023} explored the spectral signatures of Y\(^+\) and did not find this feature. However, the $4{\rm d}5{\rm p}$ to $4{\rm d}^2$ transitions were mistakenly omitted from the compiled line list in this analysis (N.~Vieira, personal communication).

\begin{figure}
    \centering
    %Zirconium/Data
    %\includegraphics[width=\linewidth,viewport=18 22 490 520 ,clip=]{Pictures/Yttrium_tardis_2.png}
    \includegraphics[width=0.99\linewidth,viewport=5 21 360 603 ,clip=]{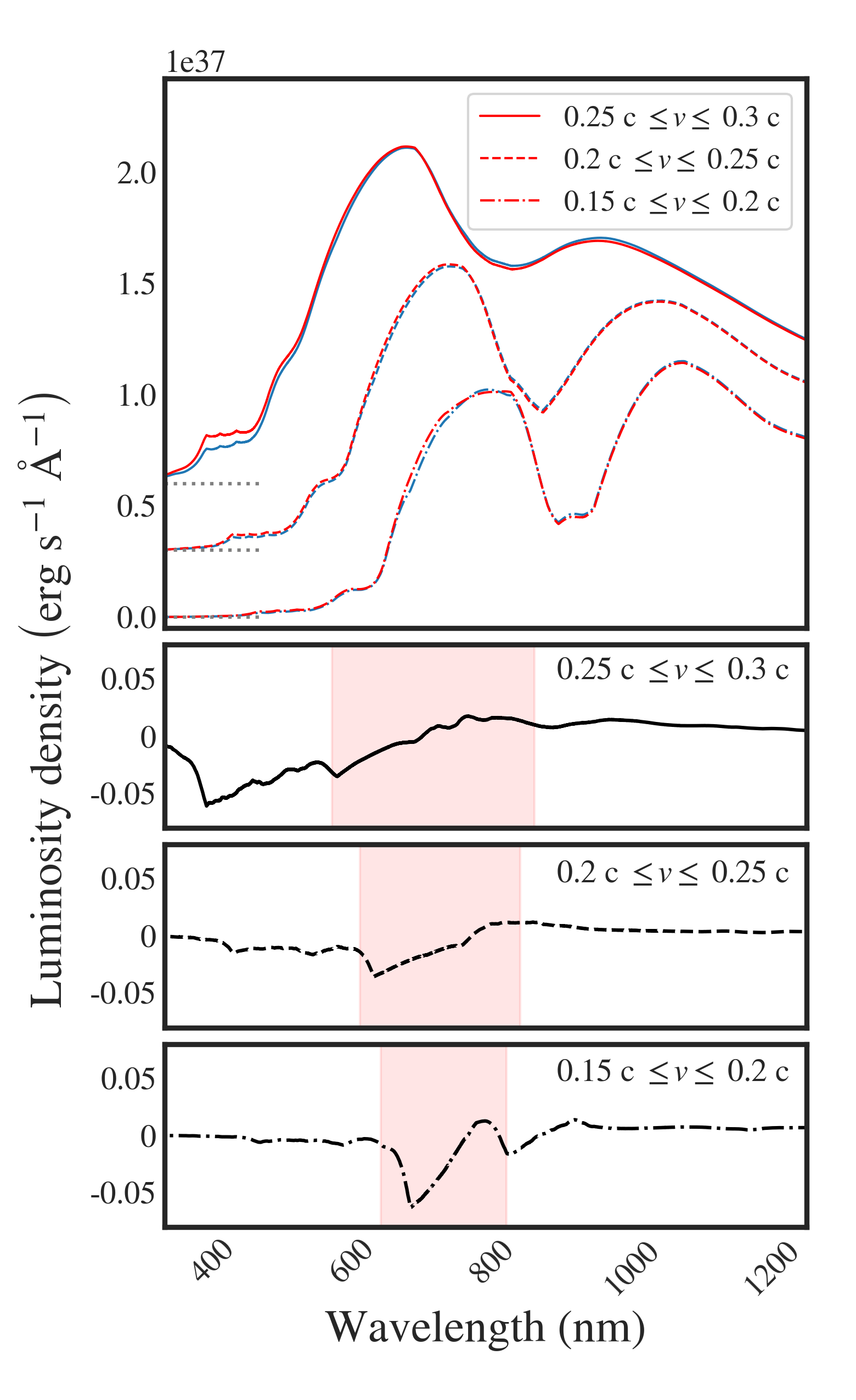}
    \caption{Example \textsc{tardis} model spectra including yttrium (blue) and excluding yttrium (red) but otherwise composed of all \rproc elements ($33 \leq Z \leq 92$) with solar \rproc abundance ratios \citep{Lodders2009,Bisterzo2014} for three different ejecta velocities. The three bottom panels show the difference in the spectral shape between the yttrium-free and yttrium-containing model for each characteristic velocity. Only when the characteristic ejecta velocity has decreased below $\sim 0.2c$ (at about 3-4 days post merger) does the P~Cygni feature become apparent, because it no longer merges with lower-wavelength yttrium features. }
    \label{fig:yttrium_velocity}
\end{figure}

\section{Discussion}\label{sec:discussion}
The 760\,nm P~Cygni feature is less prominent than its strontium P~Cygni counterpart, but shows a similar width. Furthermore, we show that it is well-fit by $4$d$5$p to $4$d$^2$ transitions in Y\(^+\). If this identification is genuine, it raises several prospects in terms of \rproc nucleosynthesis, velocity constraints, and the geometry of the ejecta.   

%\subsection{\rproc nucleosynthesis}
First, the two elements identified so far, strontium and yttrium, are both near the first \rproc-peak, which is consistent with analyses of the observed early colours of AT2017gfo \citep{Waxman2018} and arguments in \cite{Gillanders2022} and \cite{Vieira2023}, which allow a negligible to modest lanthanide abundance to reproduce the observed optical and UV flux. However, the lack of clear spectral evidence of heavier \rproc elements is somewhat surprising given the expected abundances produced in binary neutron-star mergers from typical numerical simulations \citep[e.g.][]{Ekanger2023}. This is especially notable considering the variations in abundances from different mass-ejection channels, which are believed to contribute and dominate in different epochs and depths. If both the Sr and Y line identifications are genuine, these findings reinforce the hypothesis of a light \rproc dominance despite a large radial change across epochs, which would allow for the possibility that light \rproc elements dominated the nucleosynthetic output of AT2017gfo. %In part this may be due to limited line-list for heavier element, but the  

%\subsection{Velocity and geometrical constraining power}
Second, the 760\,nm P~Cygni feature is about an order of magnitude weaker than the strontium feature in emission, though this difference is much less in absorption, while it weakens more slowly than the strontium feature at 5--6 days post-merger compared to its strontium counterpart. These differences are surprising given that Sr and Y are immediate neighbours in the periodic table and their abundance and radial distribution should be closely linked. Furthermore, the yttrium and strontium ionisation states do not vary significantly over the temperatures inferred for these epochs, with the vast majority being in the singly ionised state. However, given the significant
metastable states involved
in these transitions for both Sr and Y, it seems likely that NLTE effects change the Sr\(^+\) and Y\(^+\) populations over time as the ejecta expands. Such effects would need to be modelled accurately in order to infer the relative Sr to Y ratio. 

Importantly, a specific prediction of the yttrium identification is that the 760\,nm feature should only clearly emerge when the velocity drops to about 0.15--0.2\(c\), which is\ several days after the merger, which is precisely what observations reveal. This is not because yttrium is not in the outer layers, but because the characteristic velocities are so large that the absorption of the 760\,nm P~Cygni is coincident in wavelength with the emission peak predominantly due to the 661.4\,nm line from the $4{\rm d}5{\rm p}~{\rm a}^3{\rm P}_2$--$4{\rm d}^2~{\rm z}^3{\rm D}_3^\circ$ transition. At velocities of $0.3c,$ these features are strongly broadened and the characteristic P~Cygni feature is effectively cancelled by the summed contribution of these transitions. We illustrate this effect using a \textsc{tardis} model for three different characteristic ejecta velocities; see Fig.~\ref{fig:yttrium_velocity}. On a related note, the strong flux deficit shortward of 450\,nm in the first epoch noted in \citet{Watson2019} ---and suggested there to be related to Sr\(^+\)--- is mostly ascribed to Y\(^+\) lines in radiative transfer modelling \citep{Gillanders2022,Domoto2022,Vieira2023}. %both due to the oscillator strengths and their respective partition functions. 
A feature at $\approx760$\,nm in the first epochs could bias the inferred ejecta properties from the Sr\(^+\) lines. However, as argued in \citet{Sneppen2023}, this is unlikely to dramatically shift the constraints given the relative weakness of such features.

Lastly, we emphasise that while the statistical uncertainties are small for any specific fitting model, there are larger systematic uncertainties. This is illustrated by the relatively large variations in $v_{\rm phot}$ or $\lambda_0$ for different fitting models and epochs, respectively. For the constraints on the line shape, these systematic uncertainties may include reverberation effects, potential unmodelled blending lines, the simplicity of the assumed parameterisation of the optical depth, or even limitations in inferring the continuum flux. For the comparison with the cross-sectional velocity, $v_\perp$, the blackbody framework becomes less robust in later epochs due to emerging lines and the observed increased complexity of the spectrum. In contrast, the geometrical constraints in early epochs are derived from the more prominent Sr\(^+\) lines, which are less sensitive to the modelling of other features \citep{Sneppen2023}, and the continuum is well-fit by a blackbody \citep{Sneppen2023_bb}.
Therefore, while the Y\(^+\) feature provides an interesting approximate cross-check on the Sr\(^+\) constraints, we caution that the statistical uncertainties derived here from this 760\,nm feature do not represent the full uncertainty.

%Ideas
\section{Conclusion}
We analysed the X-shooter spectra of the kilonova AT2017gfo associated with GW170817, focussing on epochs~4 and 5, that is, 4.4 and 5.4 days post-merger, respectively. We find strong evidence for a broad P~Cygni feature with the emission component centred around 760\,nm. The velocity structure of this line is similar to that of the $\sim 1\,\mu$m line previously associated with Sr\(^+\) \citep{Watson2019}. We propose a possible identification of this feature as the sum of the 4d$^2$-4d5p transitions of Y\(^+\) between 726.42\,nm and 788.19\,nm. 
We find this to be reasonable interpretation because a \rproc abundance of yttrium yields a P Cygni feature with a central wavelength, a velocity stratification, a prominence, and a timing that match those of the observed feature.
This is the second clear identification of a particular feature with an element in a kilonova and would strengthen the case for AT2017gfo being light \rproc-element dominated.

\section{Acknowledgements}
We thank Stuart Sim and Nicholas Vieira for useful correspondence, discussions and feedback. The Cosmic Dawn Center is funded by the Danish National Research Foundation under grant number 140.  

%For the second r-process peak, there are several strong lines from Ba I and La I

% Ru I unlikely, without ionization
% Y II, 4d5p -> 4d^2, g_f = -0.57, lambda = 7881 Å

%A very prosaic discussion 

%\appendix %will make single column

\bibliographystyle{mnras}
\bibliography{refs} % if your bibtex file is called example.bib

%\section*{Appendix}
%\label{app}
%We provide the posterior distributions of key parameters for epoch~4 and 5 in Fig. \ref{fig:4posterior} and Fig. \ref{fig:5posterior} respectively. The central wavelength, $\lambda_0$ is in units of Ångstrom, while velocities are indicated in units of the speed of light, c.  

% \begin{figure}%[ht!]
%     \centering
%     \includegraphics[width=0.99\linewidth, viewport=15 15 705 725, clip=]{Pictures/750_5th_posterior_param.png}
%     \caption{Corner plot indicating posterior distribution of several key parameters from fifth epoch spectrum. The parameters indicated are 1) the cross-sectional velocity (from the blackbody normalisation fit assuming Planck cosmology), and several parameters of the 760\,nm P~Cygni line including 2) the photospheric velocity, $v_{\rm phot}$, 3) the maximum outer ejecta velocity, $v_{\rm max}$, 4) the optical depth of the line, $\tau$. Notable, the line fit is far more saturated in epoch~5, so the constraining power is far lower.}
%     \label{fig:5posterior}
% \end{figure}

\setcounter{equation}{0}
\setcounter{figure}{0}
\renewcommand{\thefigure}{A.\arabic{figure}}

\section*{Appendix A: Additional figures}
\begin{figure*}%[ht!]
    \centering    
    \includegraphics[width=0.995\linewidth]{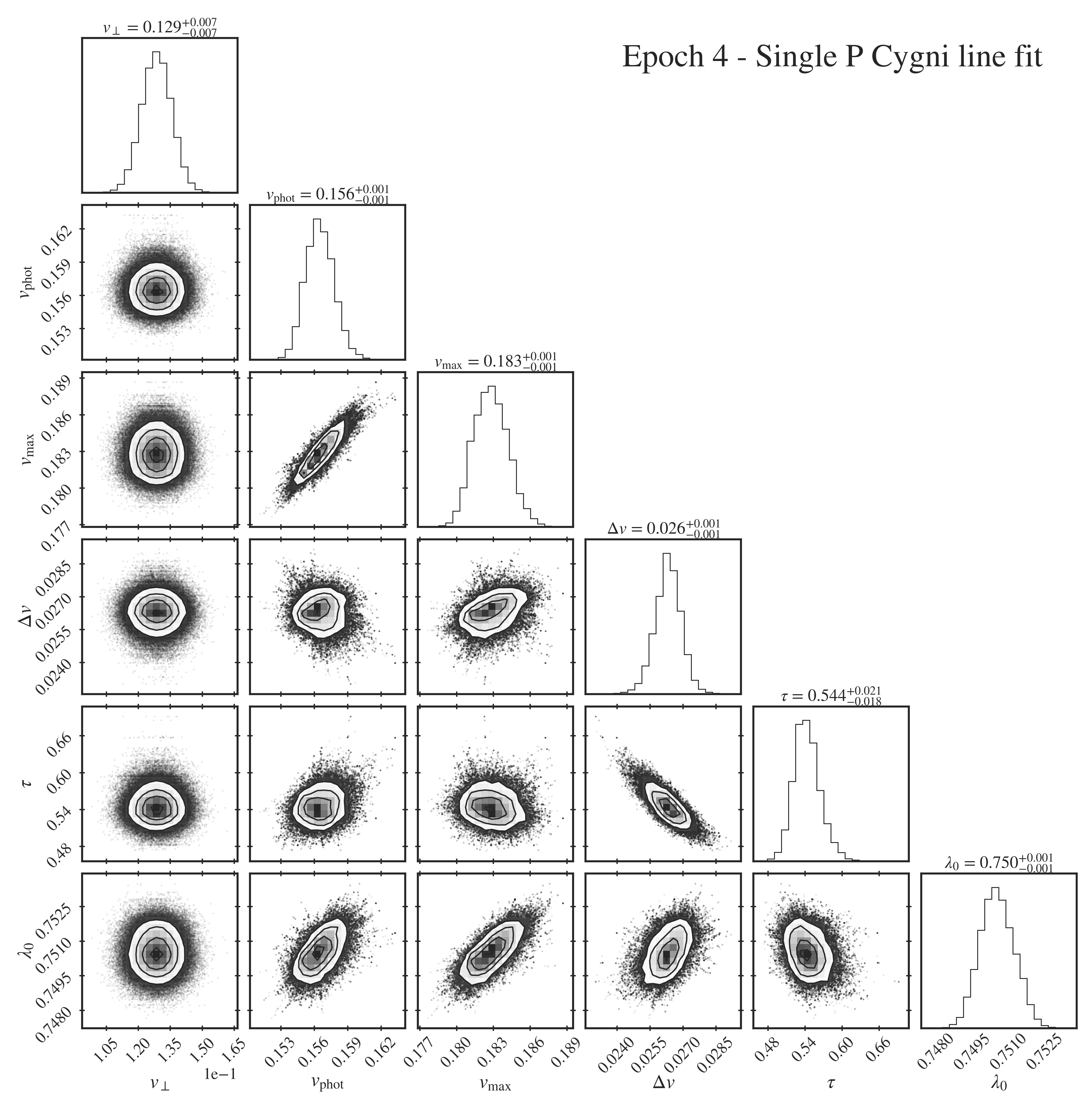}
    \caption{Corner plots showing the posterior probability distribution of key parameters for epoch~4 with a single P~Cygni line fit. %The posterior distributions have a single global minimum for epoch~4, while the fitted line is more saturated in epoch~5, so the constraining power is much lower. 
    The fit parameters are (1) the cross-sectional velocity, $v_\perp$, derived from the blackbody normalisation fit assuming Planck cosmology \citep{Planck2018}, and several parameters of the 760\,nm P~Cygni feature, including (2) the photospheric velocity, $v_{\rm phot}$, (3) the maximum outer ejecta velocity, $v_{\rm max}$, (4) the velocity range, $\Delta v = v_{\rm max}-v_{\rm phot}$, (5) the optical depth of the line, $\tau$, and (6) the central wavelength, $\lambda_0$, in $\mu$m. %(which for Epoch 5 is kept fixed at the Epoch 4 value; $\lambda = 7500$Å)
    Velocities are indicated in units of the speed of light, \(c\). These constraints show only the statistical uncertainty in the fitting model. }
    \label{fig:4posterior}
\end{figure*}

\begin{figure*}%[ht!]
    \centering    
    \includegraphics[width=0.995\linewidth]{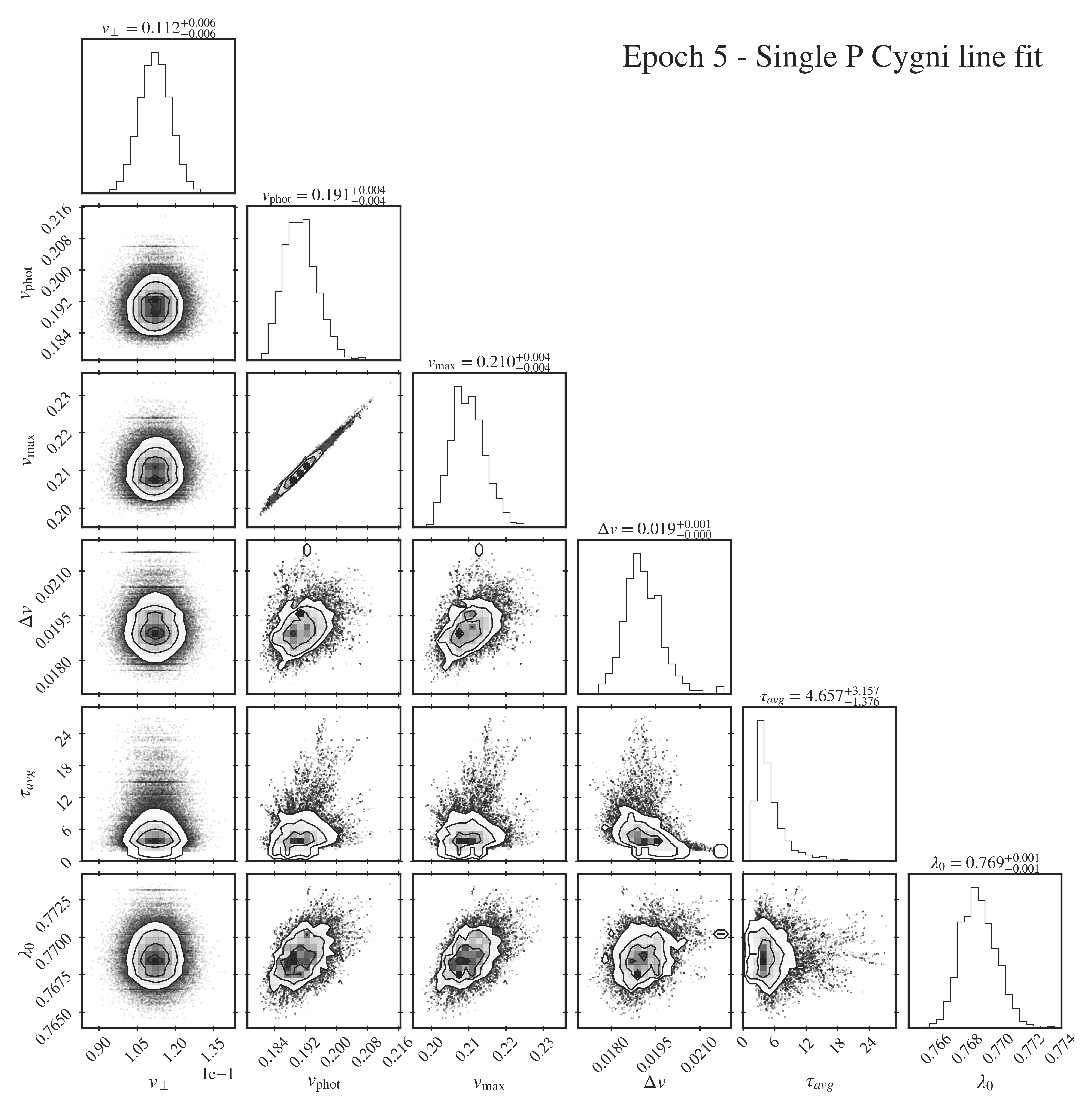}
    \caption{Corner plots showing the posterior probability distribution of key parameters for epoch~5 with a single P~Cygni line fit. %The posterior distributions have a single global minimum for epoch~4, while the fitted line is more saturated in epoch~5, so the constraining power is much lower. 
    The fit parameters are (1) the cross-sectional velocity, $v_\perp$, derived from the blackbody normalisation fit assuming Planck cosmology \citep{Planck2018}, and several parameters of the 760\,nm P~Cygni feature, including (2) the photospheric velocity, $v_{\rm phot}$, (3) the maximum outer ejecta velocity, $v_{\rm max}$, (4) the velocity range, $\Delta v = v_{\rm max}-v_{\rm phot}$, (5) the optical depth of the line, $\tau$, and (6) the central wavelength, $\lambda_0$, in $\mu$m. %(which for Epoch 5 is kept fixed at the Epoch 4 value; $\lambda = 7500$Å)
    Velocities are indicated in units of the speed of light, \(c\). These constraints show only the statistical uncertainty in the fitting model. }
    \label{fig:5posterior}
\end{figure*}

\renewcommand{\thesubsection}{\Alph{subsection}}
%\section{Appendix}
We provide the posterior distribution of the single P~Cygni line fit in Figs.~\ref{fig:4posterior} and \ref{fig:5posterior}. We emphasise that these plots show only the statistical uncertainty in the fitting model and not the (potentially sizable) systematic effects, such as those from line blending and reverberation effects.

\end{document}